\documentclass{nature}

\usepackage{graphicx}
\usepackage{siunitx}
\usepackage{amsmath}
\usepackage{amssymb}
\usepackage{color}

\title{Holographic characterisation of subwavelength particles enhanced by deep learning}

\author{Benjamin Midtvedt$^{1,2}$, Erik Ols\'en$^2$, Fredrik Eklund$^2$, Fredrik H\"{o}\"{o}k$^2$, Caroline Beck Adiels$^1$, Giovanni Volpe$^1$\& Daniel Midtvedt$^1$}
\makeatletter
\let\saved@includegraphics\includegraphics
\AtBeginDocument{\let\includegraphics\saved@includegraphics}
\renewenvironment*{figure}{\@float{figure}}{\end@float}
\makeatother
\usepackage{caption}
\begin{document}

\maketitle

\begin{affiliations}
 \item Department of Physics, University of Gothenburg, SE-41296 Gothenburg, Sweden 
 \item Department of Physics, Chalmers University of Technology, SE-41296 Gothenburg, Sweden 
\end{affiliations}

\begin{abstract}
The characterisation of the physical properties of nanoparticles in their native environment plays a central role in a wide range of fields, from nanoparticle-enhanced drug delivery to environmental nanopollution assessment.
Standard optical approaches require long trajectories of nanoparticles dispersed in a medium with known viscosity to characterise their diffusion constant and, thus, their size.
However, often only short trajectories are available, while the medium viscosity is unknown, e.g., in most biomedical applications.
In this work, we demonstrate a label-free method to quantify size and refractive index of individual subwavelength particles using two orders of magnitude shorter trajectories than required by standard methods, and without assumptions about the physicochemical properties of the medium. 
We achieve this by developing a weighted average convolutional neural network to analyse the holographic images of the particles.
As a proof of principle, we distinguish and quantify size and refractive index of silica and polystyrene particles without prior knowledge of solute viscosity or refractive index. 
As an example of an application beyond the state of the art, we demonstrate how this technique can monitor the aggregation of polystyrene nanoparticles, revealing the time-resolved dynamics of the monomer number and fractal dimension of individual subwavelength aggregates.
This technique opens new possibilities for nanoparticle characterisation with a broad range of applications from biomedicine to environmental monitoring.
\end{abstract}

Nanoparticles play a crucial role in many fields, including pharmaceutic sciences\cite{Blanco2015}, food production \cite{Singh2017, Prasad2017}, and environmental monitoring\cite{Kabir2018}. 
As particle size and composition greatly influence particle function, fast and accurate characterisation tools are essential and, ideally, should work in the native environment of the particles. 
For example, in pharmaceutic applications, the interaction between nanoparticles (e.g., protein aggregates, extracellular vesicles and viruses) and biological cells depends crucially on particle size and composition\cite{Mailander2009, Lu2009, Blanco2015, Duan2013, Zhao2011}, and studying this relation requires accurate characterisation of nanoparticles in the complex extra- and intracellular environments.
In food production, nanoparticles are used to stabilise emulsions, improving food texture and shelf life\cite{Singh2017}.
In environmental monitoring, there is a need to identify and characterise nanoparticles that enter the air, water and soil as a byproduct of industrial processes and waste disposal\cite{Kabir2018}.
In all these applications, it is often necessary to determine and monitor particle properties and activity in an environment whose physicochemical properties are unknown.

Traditionally, individual submicron particles in dispersion have been indirectly sized by analysing their diffusive Brownian motion in a solution of known viscosity (nanoparticle tracking analysis\cite{Filipe2010, Manzo2015}). 
In this approach, the trajectory of a particle is tracked, its diffusion constant $D$ is determined from the mean squared displacement of the measured trajectory, and finally, the particle radius $r$ is estimated using Stokes-Einstein relation, i.e.,  $r = {k_{\rm B} T \over 6\pi \eta D}$, where $k_{\rm B}$ is Boltzmann's constant, $T$ is the absolute temperature, and $\eta$ is the viscosity of the medium.
Even though this method is widely employed, it makes several assumptions and presents several limitations.
First, since the Brownian motion is stochastic, the particle trajectory needs to be observed over many ($>100$) time steps to achieve a reliable estimate of the particle radius, which limits the applicability of diffusion-based methods in fast processes (e.g., high-flow sorting) or when the particle sizes change dynamically (e.g., dynamic aggregation processes).
Second, the medium must be viscous with a known viscosity $\eta$, which prohibits the application of this method in biologically relevant media, which are often viscoelastic with unknown properties. 

Third, this method can only be applied to particles close to thermodynamic equilibrium because it implicitly makes use of the fluctuation-dissipation relation, excluding several processes of interest that occur out of equilibrium, e.g., in living systems and in active matter.

In order to overcome these limitations, several methods have been proposed to determine the properties of the particle by measuring its scattering properties\cite{VanderPol2014, Midtvedt2020}, instead of its motion. For example, novel microscopy techniques such as interferometric scattering microscopy (iSCAT\cite{Piliarik2014}) have made it possible to quantify the optical scattering contrast of nanoparticles close to an interface.
For particles larger than the wavelength of light, the angular distribution of the scattering intensity depends strongly on particle size, which has been exploited to characterise the size of particles in this regime\cite{Yevick2014, Wong2019}.
However, quantifying size and refractive index of subwavelength particles from optical scattering patterns in microscopy images remains challenging: The dependence of the angular distribution of the scattering on particle size is very weak, and in the limit of Rayleigh scatterers ($r \ll \lambda$) the angular distribution is independent of size. The situation is further complicated by the fact that the microscope that collects the scattering pattern acts as a low-pass filter, collecting only the light scattered in a limited range of scattering angles determined by the numerical aperture of the objective. 
Because of these problems, it has not been possible until now to use the scattering properties alone to simultaneously characterise the size and refractive index of subwavelength particles.

Here, we demonstrate that holographic imaging combined with deep learning can simultaneously characterise the size and refractive index of subwavelength particles, while using two orders of magnitude shorter trajectories than required by standard methods, and without assumptions about the physicochemical properties of the medium.
In fact, because of its potential for better, more autonomous performance\cite{cichos2020machine}, deep learning has recently been employed to successfully solve several digital microscopy problems, such as particle tracking\cite{Helgadottir2019}, anomalous diffusion characterisation\cite{bo2019measurement, granik2019single}, image segmentation\cite{Falk2019}, and image super-resolution\cite{Liu2019}.
In this work, we develop a deep-learning-powered method to determine the radius and refractive index from a series of scattering patterns of subwavelength particles recorded by an off-axis holographic microscope.

We demonstrate that this method requires two orders of magnitude fewer observations than traditional methods to reach the same accuracy in size and refractive index. 
The reduction in the number of observations required for accurate characterisation enables significantly faster characterisation of subwavelength particles, which, for example, permits us to characterise individual subwavelength particles while they are flowing in a microfluidic device. 
We also demonstrate that this method correctly characterises subwavelength particles without requiring knowledge of either particle shape or medium refractive index and viscosity, making it ideal for applications in native environments, whose physicochemical properties are often not known a priori.
Finally, we demonstrate the ability of this method to enable applications beyond the state of the art, by continuously monitoring a non-stationary process, specifically the sub-second fluctuations in size and refractive index of aggregates of nanoparticles.

\begin{figure}
\includegraphics[scale=0.85]{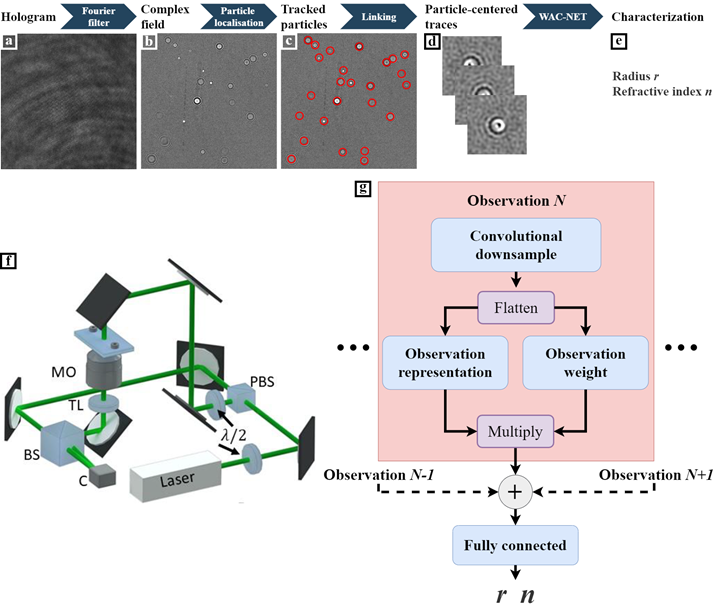} 
\caption{
\textbf{Combining holography and deep learning to characterise subwavelength particles.} 
\textbf{a}, The interference pattern is acquired with an off-axis holographic microscope and
\textbf{b}, Fourier-transformed and low-pass-filtered to produce the reconstructed field at the camera plane (here, only the imaginary part of the complex field is shown). 
\textbf{c}, These complex field images are used to track the three-dimensional particle position. 
\textbf{d}, A region of interest is selected around each detected particle for each time step.
\textbf{e}, Finally, this sequence of observations is used to characterise the radius $r$ and refractive index $n$ using a specialised neural-network design, which we call weighted average convolutional neural network (WAC-NET).
\textbf{f}, A laser beam is split into two paths by a polarising beam splitter (PBS). The relative intensities of the two paths are regulated by the initial $\lambda / 2$-waveplate. One of the beams goes through the sample, and the light scattered from the sample is collected by a microscope objective (MO), after which a tube lens (TL) provides a focused image at the camera (C). The paths are recombined using a beam splitter (BS) at a slight offset angle, resulting in an interference pattern at the camera. The particles are suspended in liquid inside a straight microfluidic chip (ChipShop). \textbf{g}, Schematic overview of WAC-NET. Each observation is downsampled and flattened using a convolutional network. From this flattened vector, a weight and a representation are calculated and multiplied. The weightings are normalised using a softmax transformation. The weighted representations are merged by summation, and the result is passed to a fully connected network that returns $r$ and $n$. }
\label{fig:1}
\end{figure}

\section*{Results}

\subsection{Experimental setup and acquisition of experimental data.}

The scattering patterns of a volume of solution containing subwavelength particles are acquired using an off-axis holographic microscope with illuminating wavelength $\lambda=\SI{633}{\nano\meter}$ and an oil-immersion objective with numerical aperture (NA) $1.3$ (see figure \ref{fig:1}f and Methods for a detailed description of the optical setup). The illuminating beam is split into two beams, one passing through the sample, and one used as a reference. Then, these beams are recombined at the camera, which records their interference pattern. 
A representative example of such a pattern is shown in figure \ref{fig:1}a. This interference pattern contains sufficient information to reconstruct the optical field at the sample with diffraction-limited resolution (figure \ref{fig:1}b) by employing a Fourier transform and a low-pass filter (see Methods for a description of the reconstruction algorithm).

The particles are imaged under constant flow in a microfluidic chip (see Methods for details on the chip). This enables us to gather scattering patterns from a large number of particles in a short amount of time (typically a few hundred particles per minute). In order to extract the scattering patterns from each particle, the particles are first localised in three dimensions using a previously published method\cite{Midtvedt2020} (figure \ref{fig:1}c) and subsequently tracked using standard methods (see Methods for a detailed description). A region of $64\times64$ pixels centred around each particle detection is extracted. In this way, we record and store a collection of scattering patterns of each particle (figure \ref{fig:1}d), which is subsequently analysed using a neural network (figure \ref{fig:1}e).

\subsection{Structure of the neural network.}

Since the particles are smaller than the wavelength of the illuminating light, the features that are available for determining particle size are contained in the high-frequencies of the image spectrum (corresponding to high scattering angles)\cite{Novotny2006}. 
At low signal-to-noise ratio, it is prohibitively difficult to analytically fit these features and determine the particle properties.
Thus, to achieve a good signal-to-noise ratio, it is necessary to  average over multiple observations of the same particle. However, direct averaging of observations requires perfect three-dimensional localisation, which is in practice impossible to achieve because of the error intrinsic to any localisation technique. 

In order to overcome this problem, we employ machine learning.
While in standard methods the user explicitly defines a set of rules to convert the input data to the desired output, in machine learning the rules are inferred by providing the machine-learning model with a large collection of input--output pairs --- this process is known as training of the machine-learning model\cite{cichos2020machine}. Neural networks are a very successful subset of machine learning models\cite{chollet2017python}, and consist of interconnected layers which apply some, often simple, transformation to their input and pass the result to some other layer. The transformation each layer applies can be controlled by weights, which are optimised during the training. 

The architecture of the neural network we employ is shown in figure \ref{fig:1}g. We have optimised this architecture to overcome the two main challenges that we face when characterising size and refractive index of subwavelength particles. First, the neural network needs to be able to use all available observations of a particle to refine its prediction. 
Second, it needs to be able to account for the fact that all the individual observations may not be equally representative (due to, e.g., uncorrelated noise or interference of the scattering patterns of nearby particles). This prevents us from averaging at the output, while direct averaging at the input is prohibited by imperfect centring of the scattering patterns, as explained above. In order to overcome these challenges, we develop a network architecture which performs a weighted average of the observations at an intermediate step, between input and output. We achieve this by first letting the network transform each complex scattering pattern into a one-dimensional vector and a single scalar representing the confidence in the obtained vector representation, using a series of convolutional layers and max-pooling layers. The network then performs a weighted averaging of these vectors using the confidence values as weights, before predicting the size and refractive index of the particle by applying a series of fully-connected layers to the averaged vector. We call this neural-network architecture weighted average convolutional neural network (WAC-NET).

\begin{figure}
\centering
\includegraphics[scale=.9]{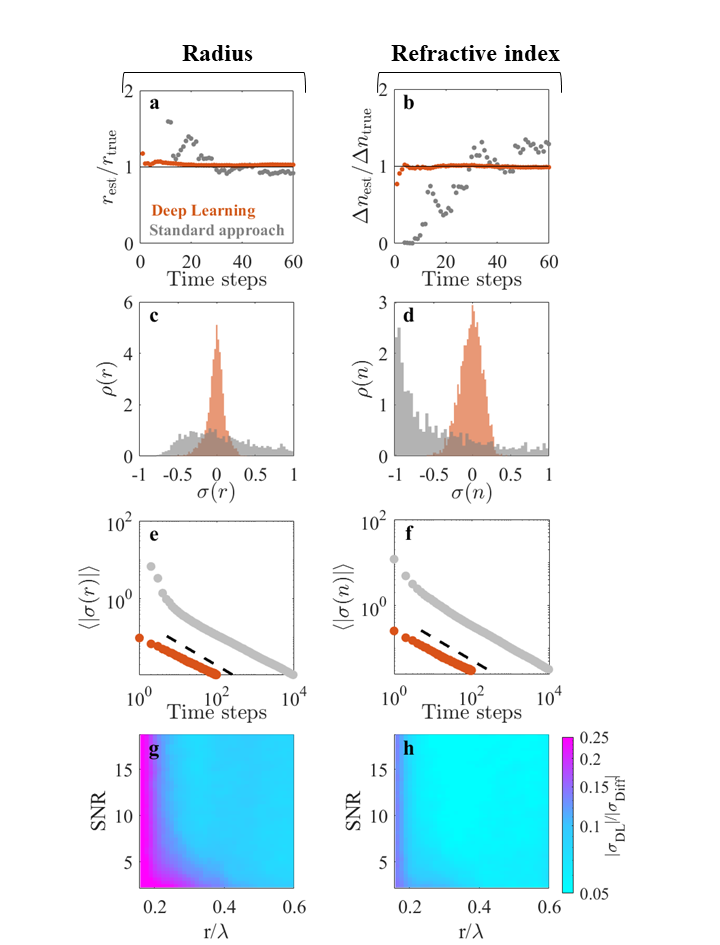} 
\end{figure}
\begin{figure}
\caption{
\textbf{Deep learning enhances particle characterisation on simulated data.} 
\textbf{a}-\textbf{b}, Radius $r_{\rm est}$ normalised by the ground-truth radius $r_{\rm true}=\SI{0.20}{\micro\meter}$ and refractive index difference $\Delta n_{\rm est}$ normalised  by the ground-truth refractive index difference $\Delta n_{\rm true}=0.25$; $r_{\rm est}$ and $\Delta n_{\rm est}$ are estimated using deep learning (orange symbols) and a conventional diffusion-based approach (grey symbols) as a function of the number of observations of the particle. The deep-learning approach requires fewer observations to converge to the ground-truth values.
\textbf{c}-\textbf{d}, Distributions of the relative errors $\sigma_{r}$ and $\sigma_{n}$ for $r_{\rm est}$ and $n_{\rm est}$, respectively, on the characterisation of 2000 simulated particles using 5 observations.
While the standard approach (grey histograms) fails to characterise the particles, the deep-learning approach (orange histograms) characterises the radius to within $5\%$ and refractive index to within $10\%$ (mean absolute errors). 
\textbf{e}-\textbf{f}, Scaling of the absolute relative errors $\langle|\sigma_{r}|\rangle$ and $\langle|\sigma_{n}|\rangle$ as a function of the number of observations used for the characterisation. The dashed line represent the scaling $1/\sqrt{N}$, where $N$ is the number of observations.
\textbf{g}-\textbf{h}, The ratio of the absolute relative error made by the deep-learning approach ($\sigma_{r, \rm DL}$) and the diffusion-based approach ($\sigma_{\rm Diff}$) using 20 observations shows that the deep-learning approach provides an order of magnitude more accurate characterisation both for particle radius (\textbf{g}) and refractive index (\textbf{h}) in a wide range of particle radii $r/\lambda$ (where $\lambda$ is the wavelength of the illuminating light), even on noisy images (signal-to-noise ratio $\rm{SNR} < 10$). 
}
\label{fig:2}
\end{figure}

\subsection{Training of the neural network.}

In order to train the WAC-NET, we first must generate a training set that is representative of the data produced by the optical setup. Using experimental data for training the WAC-NET is challenging, because they are limited in their number and have intrinsic experimental errors. We therefore generate synthetic training data by simulating the complex scattering patterns of subwavelength particles using Mie theory (see Methods for details on the scattering pattern simulation). The scattering patterns are convolved with the optical transfer function of the microscope, determined by analysis of reference particles of known size and refractive index (see Methods for details on this calibration procedure). 
We train the WAC-NET to infer particle sizes and refractive indices within a broad range of values by simulating $10^6$ scattering patterns cropped to 64-pixel-by-64-pixel images from particles having radii uniformly distributed in the range \SI{115}{\nano\meter}$\leq r\leq$\SI{500}{\nano\meter} and refractive indices uniformly distributed in the range $1.36\leq n \leq 1.9$, dispersed in a medium with refractive index $1.33$. In order to make the training set representative of the experimental data, the observations of each particle are corrupted with inaccurate three-dimensional centring of the particle (with a standard deviation of the error of \SI{0.1}{\micro\meter}, similar to the size of individual pixels) and synthetic noise in the form of spatially correlated background noise (with amplitude determined from empty slices of the image). A random number between 5 and 50 such observation are stacked to form the synthetic representation of a traced particle.
An additional advantage of using synthetic data is that new data can be continuously generated during the training, eliminating the risk of overfitting due to a limited training set.

\subsection{Neural network performance on simulated data.}

We first test the trained WAC-NET on simulated data.
We simulate the scattering patterns of a set of 2000 stacks of \SI{0.20}{\micro\meter} (radius) polystyrene (PS) spheres (refractive index 1.58, illumination wavelength $\lambda=633$\SI{}{\nano\meter} and objective ${\rm NA}=1.3$) dispersed in water, each stack consisting of 100 observations of a single particle.
The orange symbols in figures \ref{fig:2}a and \ref{fig:2}b represent the estimated values of the radius $r_{\rm est}$ and refractive index $n_{\rm est}$ as a function of the number of time steps during which a representative particle is observed. 
It can be seen that these estimates converge to the ground truth values (solid black lines) within only a few observations.
For comparison, the characterisation based on conventional particle tracking (grey symbols) produces significantly worse results than those obtained by the WAC-NET and do not converge to the ground-truth value even when using 60 time steps.

To compare the network performance with regard to size determination to particle sizing using conventional particle tracking measurements, we simulate 2000 Brownian motions, each having $N = 10^4$ time steps with time step $\Delta t$. To estimate the  diffusion constant corresponding to a Brownian motion, we use the standard approach based on the mean squared displacement in a time step, i.e., $D={1\over N}\sum_{i=1}^{N}{\Delta x_i^2 \over \Delta t}$, where $\Delta x_i$ is the length of the $i
^{\rm th}$ displacement. 
To determine the refractive index, there is no standard method for subwavelength particles. Nonetheless, it has previously been demonstrated that measuring either scattering intensity\cite{VanderPol2014} or phase contrast\cite{Midtvedt2020} in combination with analysis of the Brownian motion allows simultaneous determination of size and refractive index\cite{Midtvedt2020}. In order to compare the performance of WAC-NET with existing methods for the determination of refractive index, we assume that the phase contrast of the particles are known exactly at each time step and estimate the refractive index as described in Ref. \citenum{Midtvedt2020}.

Figures \ref{fig:2}c and \ref{fig:2}d show the performance of WAC-NET (orange histograms) and diffusion-based characterisation (grey histograms) on the ensemble of 2000 particles using 5 time steps for characterisation. While the distributions of estimates produced by WAC-NET are well-defined and centred around the true value, the diffusion-based estimates are scattered in a wide range and are not centred around the true value, demonstrating that WAC-NET performs significantly better than diffusion-based methods when using few time steps.
These results highlight the difference in the underlying principle of the two approaches. In contrast to the stochastic nature of the diffusional motion, the scattered field is deterministic, allowing our deep learning approach to directly infer particle size from the optical scattering properties of the particle. Repeated observations of the same particle are used to refine this inference. Consequently, it is possible to reach high accuracy in particle sizing using only a few observations. 

To determine whether the accuracy achieved by the WAC-NET is better than that of the diffusion-based method also when using many time steps for the characterisation, we study the scaling of the estimated error of the two approaches (defined as $\sigma(y)=\langle (y_{\rm pred}/y_{\rm true}-1)\rangle$), where $y$ represents either size or refractive index, with the number of time steps in figures \ref{fig:2}e and \ref{fig:2}f. We find that the error of both methods decreases as $1/\sqrt{N}$, where $N$ is the number of time steps (dashed black lines), with the WAC-NET consistently producing more than an order of magnitude better results than the diffusion-based method. Consequently, the error of both methods can be parametrised as $\sigma(y)=\sigma_0/\sqrt{N}$, where $\sigma_0$ is a proportionality constant. In the case of diffusional motion, the determination of the diffusion constant in essence amounts to estimating the variance of the distribution of particle displacements. Assuming that $N$ displacements are known exactly, the uncertainty in the estimated variance of the underlying distribution is given by $\sigma^2=2/N$. This provides a lower bound on the proportionality constant for diffusional motion that cannot be improved upon by changes to the experimental parameters\cite{Vestergaard2014}. In the case of the deep learning approach, the prefactor is instead determined by the particle type and the quality of the image. 

Figures \ref{fig:2}g and \ref{fig:2}h plot the prefactor $\sigma_0$ as a function of the signal-to-noise ratio (SNR) of the image and the ratio of the particle radius to the illuminating wavelength ($r/\lambda$).
This allows us to compare the performance of the network to that of the diffusion-based approach for a wide range of particle sizes and noise levels. We find that the deep-learning-based particle characterisation method is more than one order of magnitude more accurate than diffusion-based methods for a fixed number of observations in a broad range of particle sizes, extending down to $r\approx \lambda/3$. Owing to the scaling of the errors of the two methods with the number of observations, the diffusion-based method will need more than two orders of magnitude more observations to reach the same accuracy in this region of particle sizes. Furthermore, the network shows a considerable increase in accuracy compared to a diffusion-based approach even in poor illumination conditions (down to $\mathrm{SNR}=2$). Also, note that the positions of the simulated particles were known exactly, while, in any experimental setting, error in the particle localisation inevitably induces additional uncertainties in the determination of the diffusion constant; therefore, the stated improvement in accuracy is a lower bound.

\begin{figure}
\centering
\includegraphics[scale=0.7]{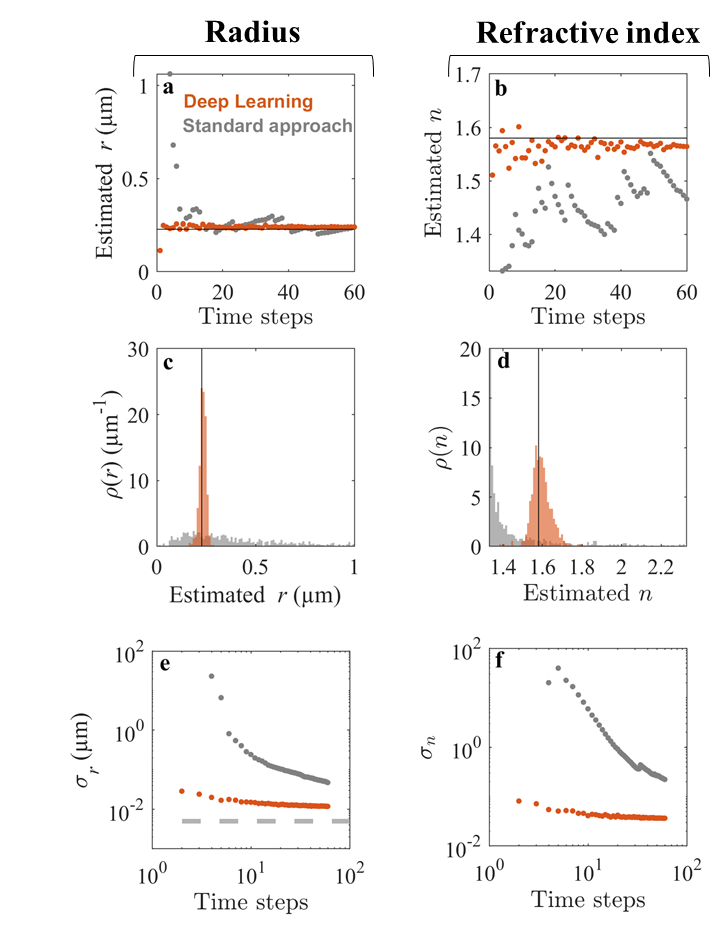} 
\end{figure}
\begin{figure}

\caption{
\textbf{Deep learning enhances particle characterisation on experimental data.} 
\textbf{a}-\textbf{b}, 
Estimated radius $r$ and refractive index $n$ of a polystyrene sphere as a function of the number of observations using deep learning (orange symbols) and a standard diffusion-based method (grey symbols). The manufacturer-provided nominal radius is 0.228$\pm$\SI{0.005}{\micro\meter} and the bulk value for the polystyrene refractive index is between 1.55 and 1.59.
The accuracy of the deep-learning approach surpasses the diffusion-based approach with all numbers of observations. 
\textbf{c}-\textbf{d}, Distribution of $r$ and $n$ on an experimental sample of 256 particles using 5 observations.
While the diffusion-based approach (grey histograms) fails to characterise either property, the deep-learning approach (orange histograms) provides estimates of the radius with $\pm$\SI{16}{\nano\meter} standard deviation and of the refractive index with $\pm 0.05$ standard deviation. 
\textbf{e}-\textbf{f}, The scaling of the standard deviations $\sigma_{r}$ and $\sigma_{n}$ of the measured $r$ and $n$ of the population as a function of the number of observations. The standard deviation of the deep-learning approach (orange symbols) reaches $\pm$\SI{9}{\nano\meter} in radius and $\pm 0.03$ in refractive index units using 60 observations. This performance is much better than that of the standard diffusion-based approach (grey symbols): $\sigma_{r}$ is comparable to the nominal size variation stated by the manufacturer ($\pm$\SI{5}{\nano\meter} standard deviation, grey dashed line in \textbf{e}) and $\sigma_{n}$ is within the range of variability of the polystyrene refractive index.}
\label{fig:3}
\end{figure}

\subsection{Neural network performance on experimental data.}

To verify that the increase in accuracy of the deep learning characterisation transfers to experimental data, we recorded holographic videos of \SI{0.23}{\micro\meter} radius polystyrene spheres flowing through a microfluidic chip (see Methods for further details). 
In figures \ref{fig:3}a and \ref{fig:3}b, the characterisation of a representative particle is shown as a function of the number of time steps used for characterisation. We find that the WAC-NET converges to estimates close to the expected values of size and refractive index using only a few time steps (orange symbols). The characterisation based on conventional particle tracking instead shows considerably larger fluctuations (grey symbols). Similarly, WAC-NET produces predictions consistent with the expected values on the ensemble of particles using only five time steps (orange histograms in figures \ref{fig:3}c and \ref{fig:3}d, expected values of size and refractive index are shown as solid black lines). On the other hand, the diffusion-based approach does not correctly characterise the sample using only five time steps (grey histograms).

When considering the scaling of the error of the predictions with the number of time steps, we find that the accuracy of the deep-learning approach, when applied to experimental data, appears to saturate around a standard deviation of \SI{9}{\nano\meter} and $0.03$ refractive index units for the size and refractive index, respectively (figures \ref{fig:3}e and \ref{fig:3}f). This saturation partly reflects the distribution of sizes and refractive indices within the sample. The variation in size of these particles was given as $\pm$\SI{5}{\nano\meter} standard deviation by the manufacturer. The remaining variation stems from noise not averaged by multiple observations (e.g., long-range temporal correlations in the image noise and spatially variable aberrations not perfectly accounted for). Since the particles are imaged under flow, they could not be tracked for sufficiently long to determine a similar saturation point for the conventional particle tracking approach.

\subsection{Characterisation of a multicomponent sample.}

\begin{figure}
\hspace*{-2cm} 
\includegraphics[scale=0.5]{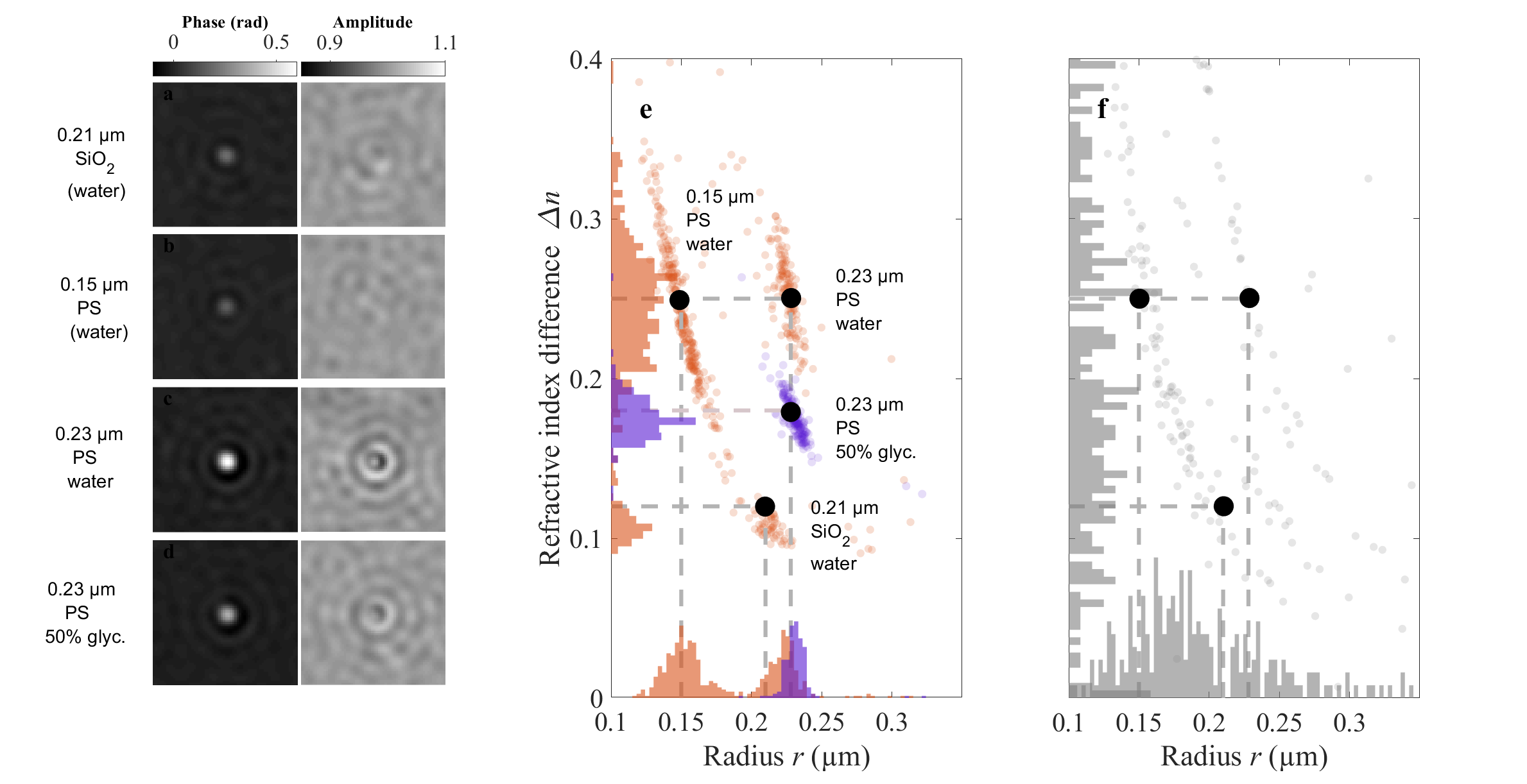} 
\caption{
\textbf{Deep-learning approach performance on particle mixtures and in different environments.}
\textbf{a}-\textbf{d}, Phase and amplitude signals from a representative particle from each characterised population. \textbf{e}, The deep-learning approach (using 60 observations) distinguishes and correctly characterises subpopulations in a multicomponent mixture dispersed in water, consisting of \SI{0.21}{\micro\meter} silica (${\rm SiO_2}$), \SI{0.15}{\micro\meter} polystyrene (PS) and \SI{0.23}{\micro\meter} PS particles.
Furthermore, the deep-learning approach accurately characterises the radius $r$ and the refractive index difference $\Delta n$ of \SI{0.23}{\micro\meter} PS particles dispersed in a 50\% glycerol/water mixture, demonstrating that the measurements do not rely on detailed knowledge of the properties of the solution. The intersections of the dashed lines represent the expected positions of the populations. \textbf{f}, When using a diffusion-based approach to characterise the multicomponent sample the size and refractive index distributions become very broad, and the two smallest subpopulations cannot be distinguished when using 60 observations for the characterisation.}
\label{fig:4}
\end{figure}

As a test of the performance of the network in a more complex and realistic scenario, we analyse a multicomponent sample consisting of polystyrene (PS) particles of two different sizes (modal radii \SI{0.23}{\micro\meter} and \SI{0.15}{\micro\meter}, refractive index 1.58) and silica particles (modal radius \SI{0.21}{\micro\meter}, refractive index $1.45$). This sample is challenging to analyse using standard methods since \SI{0.23}{\micro\meter} polystyrene and \SI{0.21}{\micro\meter} silica are similar in size and are therefore hard to distinguish by diffusion alone, especially when only a few observations are available. In addition, \SI{0.15}{\micro\meter} PS and \SI{0.21}{\micro\meter} silica induce similar phase contrasts, and their scattering patterns are consequently visually similar. This is highlighted in figures \ref{fig:4}a-c, where the phase and amplitude contrasts of characteristic particles within the three subpopulations are shown.

Despite these similarities, the deep-learning approach successfully distinguishes the three subpopulations and accurately determines their modal characteristics (figure \ref{fig:4}e). Specifically, the sizes and refractive indices are estimated by WAC-NET to be $r=0.15\pm 0.01$\SI{}{\micro\meter} (standard deviation), $n=1.58\pm0.06$  (\SI{0.15}{\micro\meter} PS); $r=0.22\pm 0.02$\SI{}{\micro\meter}, $n=1.44\pm 0.02$ (\SI{0.21}{\micro\meter} silica); and $r=0.22\pm 0.01$\SI{}{\micro\meter}, $n=1.58\pm 0.03$  (\SI{0.23}{\micro\meter} PS). The slightly larger variation observed in the smallest subpopulation is consistent with the discussion above in connection with figures \ref{fig:2}g and \ref{fig:2}h: in our setup, a radius of \SI{0.15}{\micro\meter} corresponds to a ratio $r/\lambda\approx 0.24$, which is within the region where the precision of the WAC-NET starts to decay. In order to understand the correlation between estimated size and refractive index in this population, we note that the scattering amplitude for particles in this size regime scales approximately with $V\cdot\frac{n^2-n_{\rm m}^2}{n^2+2n_{\rm m}^2}\propto V\cdot \Delta n/n_m$ to lowest order in $\Delta n = n-n_{\rm m}$, and where $n$ and $n_{\rm m}$ are the refractive index of the particle and the medium, respectively\cite{Midtvedt2020, Novotny2006}. For particles with $r\lesssim \lambda/4$, the WAC-NET can still predict the scattering amplitude via the product $V\cdot \Delta n$ (see supplementary figure 1a), but the precision by which the network can separate size and refractive index from this product deteriorates somewhat in this size regime. As a consequence, for particles with $r\lesssim \lambda/4$ the estimated sizes and refractive indices are correlated in such a way as to keep the product $V\cdot \Delta n$ constant (supplementary figure 1b).

In contrast, the approach based on conventional particle tracking fails to distinguish \SI{0.15}{\micro\meter} PS and \SI{0.21}{\micro\meter} silica from each other (figure 4f). Thanks to the difference in phase contrast, the \SI{0.23}{\micro\meter} PS population can be distinguished, but the distribution of estimated size and refractive index becomes very broad.

\subsection{Characterisation of particles in an unknown environments.}

To test the performance of the network in situations where the properties of the solute are not known a priori, we also record holographic videos of \SI{0.23}{\micro\meter} polystyrene particles dispersed in a 50\% glycerol/water mixture. The presence of glycerol changes both the viscosity and the refractive index of the solution, and thus this solute is qualitatively different from the solute used for network calibration. In the case of holographic imaging, a change in the refractive index of the environment primarily changes the scattering amplitude but leaves the spatial structure of the scattering pattern approximately unchanged (figure \ref{fig:4}d, cfr. figure \ref{fig:4}c). Thus, we expect the network to correctly recognise the particles having the same sizes in both environments but with a different relative refractive index. Indeed, we found that the particle populations measured in the two solutions overlap in size, but are shifted in refractive index (figure \ref{fig:4}e). In order to relate the shift in estimated refractive index of PS particles to the difference between the refractive index of water ($n_{\rm w}=1.33$) and that of the glycerol/water mixture ($n_{\rm g/w}\approx 1.40$), we note that the shift in scattering amplitude is proportional to $n_{\rm g/w}-n_{\rm w}$, valid to lowest order in $n_{\rm g/w}-n_{\rm w}$. Thus, the shift in scattering amplitude is expected to produce a shift in the estimated refractive index of the particles that correspond to the difference in refractive index between the solutions. Indeed, we find that this is the case, demonstrating the capacity of the deep-learning-based method to characterise particles without prior knowledge of the physicochemical properties of their surrounding environment.

\subsection{Monitoring clusters of nanoparticles.}

\begin{figure}
\centering
\includegraphics[trim={2.2cm 0 2.6cm 0},clip,scale=0.61]{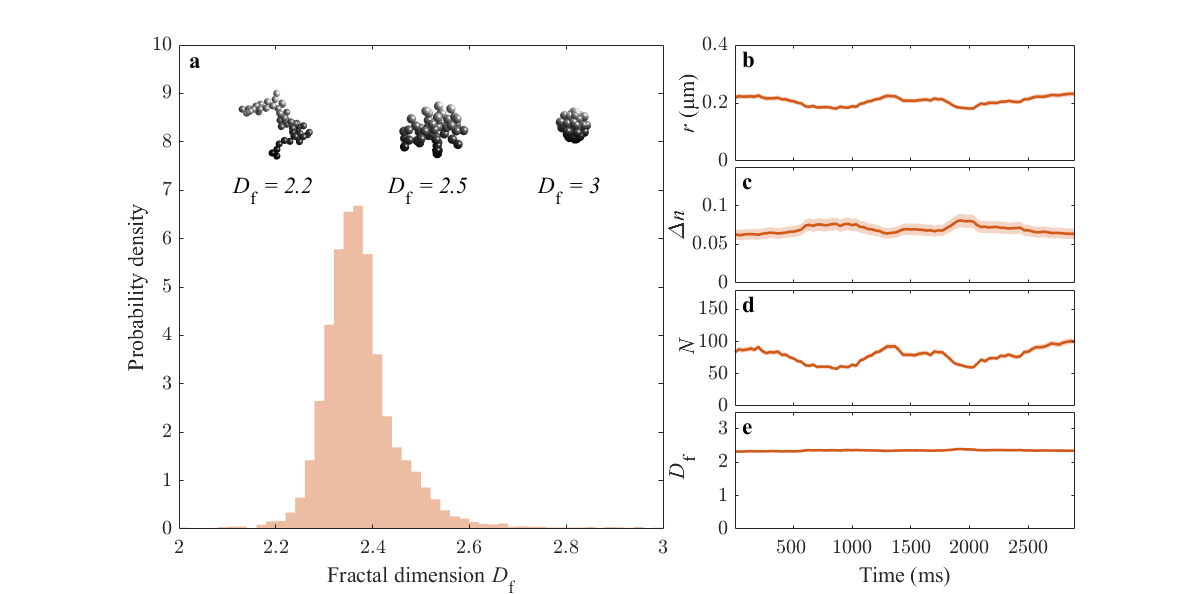} 
\caption{
\textbf{Time-resolved dynamics of nanoparticle clusters.}
\textbf{a}, The ensemble of particle clusters formed by \SI{31}{\nano\meter} radius polystyrene (PS) monomers features an average fractal dimension $D_{\rm f}$ close to $2.35$.
The insets on the top show some pictorial depictions of possible clusters with various fractal dimensions.
\textbf{b}-\textbf{e}, Time-resolved behaviour of a representative cluster, characterised in terms of its radius $r$ (\textbf{b}), refractive index difference $\Delta n$ (\textbf{c}), number of monomers $N$ (\textbf{d}), and fractal dimension $D_{\rm f}$ (\textbf{e}). While $r$, $\Delta n$, and $N$ greatly vary over time, $D_{\rm f}$ remains stable in time.
The shaded regions represent the estimated standard deviation of the error. 
The cluster is characterised using a moving window of 20 observations, acquired at a frame rate of 30 frames per second. The fractal dimension of the cluster is estimated based on the scaling of its size and refractive index with the number of monomers, assuming a known monomer with radius  \SI{31}{\nano\meter} and refractive index 1.58.
}
\label{fig:5}
\end{figure}

In the experiments presented until now, we have evaluated the network performance on spherical particles whose properties do not change over time. In order to test it in a more dynamic scenario beyond the state of the art of what can be done with standard techniques, we image a sample consisting of a solution of \SI{31}{nm} radius polystyrene nanoparticles freely diffusing in a microfluidic chip and forming dynamic clusters  (nanoparticle clustering is induced by adding a droplet ($\sim$\SI{20}{\micro\liter}) of saturated NaCl solution to the inlet of the microfluidic chip, which alters the Debye screening length and destabilizes the suspension). These clusters are in general non-spherical, and their size and refractive index fluctuate over time. Since clustering of nanoparticles affects their reactivity, understanding how such clusters form, how they interact, and how their size and morphology evolve in time is important to understand and predict their behaviour and performance\cite{Hotze2010}. For instance, determining whether the formation of clusters is irreversible (aggregation) or reversible (agglomeration) is challenging, presently requiring combining a multitude of measurement techniques\cite{Sokolov2015}. In fact, with the exception of liquid-cell transmission electron microscopy, which suffers from complicated sample preparation and low throughput, existing methods for characterising nanoparticle clusters only provide snapshots of the aggregation/agglomeration process, and thus cannot be used to temporally monitor the clusters\cite{Liu2016}. 

A signature of cluster formation is the fractal nature of the resulting structure. A fractal cluster can be characterised by its fractal dimension, $D_{\rm f}$, which dictates the scaling of the size with the number of monomers, $N$, in the cluster (see insets in figure \ref{fig:5}a): $(r/r_0) = k \cdot N^{1/D_{\rm f}}$, where $r$ is the radius of the cluster, $r_0$ is the radius of the monomers and $k$ is a proportionality constant ($k\approx 1$ for clusters with fractal dimensions $D_{\rm f}>2$,\cite{Fung2019} so in the following we set $k=1$) . For close-packed spheres, the radius of the cluster scales as $(r/r_0)\sim N^{1/3}$. 
The effective refractive index $n$ of a cluster formed by identical spheres with radius $r_0$ and refractive index $n_0$ follows the Maxwell-Garnett relation, $L(n)=(r_0/r)^3 N \cdot L(n_{\rm 0})=(r_0/r)^{3-D_f}L(n_{\rm 0})$, with $L(n) = \frac{n^2-n_m^2}{n^2+2n_m^2}$, with $n_{\rm m}$ being the refractive index of the surrounding medium\cite{Wang2016}. The latter expression can also be used to directly estimate the fractal dimension of a cluster, given that the radii and refractive indices of both the monomers and the cluster as a whole are known. 

For diffusion-limited clustering of monodisperse monomers, the fractal dimension is expected to be around $D_{\rm f}=2.5$, whereas the value is lowered to $D_{\rm f}=1.75$ if cluster-cluster clustering is taken into account\cite{Schaefer1984}. 
In line with this, previous measurements of the fractal dimension of colloidal clusters have yielded values in the range $D_{\rm f} \sim 1.6-2.3$, with the exact value depending on multiple factors, such as the concentration of salt in the solution and the concentration of monomers \cite{Schaefer1984,Zhou1991,Asnaghi1992,Carpineti1990,Amal1990}. Furthermore, the fractal dimension has been predicted to decrease with time, with early clusters having $D_{\rm f}\sim 2.5$, decreasing to $D_{\rm f}\sim 1.8$ as a function of time\cite{Kostoglou2001}. Based on these results, we expect the fractal dimension in our experiment to be close to $D_f=2.5$. Consistent with this, we find that the fractal dimension of the structures are $D_f=2.35 \pm 0.1$ (figure \ref{fig:5}a).
This demonstrates that the deep-learning-based approach is sufficiently accurate even on non-spherical subwavelength particles to estimate their fractal dimension, and that the method does not require the shape of the particles to be known a priori. 

The characterisation of the clusters using the WAC-NET permits us also to investigate the temporal dynamics of the properties of individual nanoparticle clusters. To do so, we characterise the clusters using a sliding window of 20 time steps (at a frame rate of 30 fps), providing subsecond temporal resolution of size and refractive index of individual clusters. Fluctuations in these properties can, in general, be of two different physical origins: association/disassociation of monomers to the cluster will change the number of monomers in the cluster, without changing the fractal dimension, whereas internal rearrangements of monomers will cause a change to the fractal dimension, at a fixed monomer number. Thus, by monitoring the number of monomers $N$ and fractal dimension $D_{\rm f}$, the physical origin of fluctuations in size and refractive index can be determined. Figures 5b and 5c show the behaviour of a representative cluster, displaying fluctuations of both size and refractive index. These fluctuations are attributed to a fluctuation in the number of monomers in the cluster (figure 5d). Strikingly, the estimated fractal dimension of the structure remains constant in time despite the large temporal size fluctuations (figure 5e), indicating that internal rearrangement of monomers does not occur at the time scale of the measurement. The large fluctuations of the number of monomers in the clusters can only be explained by continuous association/disassociation of monomers to the clusters, strongly suggesting that the clusters are in fact agglomerates (weakly/reversibly bound), rather than aggregates (strongly/irreversibly bound).

\section*{Discussion}

We have demonstrated the potential for deep-learning-enhanced optical characterisation of subwavelength particles. 
Going beyond standard approaches, we have shown that the scattering patterns of individual particles recorded in off-axis holographic imaging contain sufficient information to extract size and refractive index of dielectric particles of radius down to $r=$ \SI{0.15}{\micro\meter}.

With standard approaches, individual characterisation of dispersed particles in the subwavelength regime requires the analysis of the Brownian motion of the particles through particle tracking. By instead using the optical scattering pattern to deduce size and refractive index, our approach offers several advantages compared to traditional approaches. Specifically, our method does not require prior knowledge of the physical properties of the surrounding medium, such as its viscosity or refractive index. This is particularly important in industrial and biological systems, where particles need to be characterised in complex environments. Furthermore, our method provides temporally resolved measurements of size and refractive index of the particles, potentially enabling direct monitoring of interaction kinetics of particles in their native environment.

Taking the approach presented in this article, we achieve an estimated error in particle sizing of $<10\%$ using as few as 5 time steps, orders of magnitude faster than diffusion-based methods. In addition to particle size, our method also estimates the particle refractive index with comparable accuracy, serving as a proxy for particle composition. The refractive index and particle size are estimated from the optical scattering profile, with no reference to particle motion, thus enabling accurate characterisation in arbitrary environments. 

It is also worth noting that the approach we propose requires only a few observations of each particle to accurately characterise both size and refractive index. Thus, deep-learning-based characterisation enables temporally resolving dynamical changes in size and refractive index of individual particles on subsecond time scales. We demonstrate this capability by monitoring the aggregation kinetics of a sample consisting of polystyrene nanoparticles in a high-salt environment. Despite the unknown geometrical shape of the scatterers, our method can resolve changes in the number of monomers as well as the resulting changes in aggregate size and refractive index. This characterisation is sufficiently accurate to provide a reliable time-resolved estimate of the fractal dimension of individual aggregates, which, to our knowledge, has not previously been reported.

Taken together, our results show that deep-learning-enhanced analysis of holographic scattering patterns allow improved accuracy in particle size and refractive index determination by more than an order of magnitude compared to traditional methods. The characterisation is performed without assumptions on the physical properties of the environment and shape of the particle and can be performed with subsecond temporal resolution. This opens up the possibility to temporally resolve size and composition of individual subwavelength particles in their native environment. We anticipate that the drastic improvement in single-particle characterisation offered by this technique will find widespread application in any area where subwavelength particles play an important role, ranging from industrial processes to drug discovery and pharmaceutics.

\begin{methods}

\subsection{Particles and chemicals.}
The used monodisperse particles are \SI{0.031}{\micro\meter} (modal radius) and \SI{0.15}{\micro\meter} (modal radius) polystyrene (Invitrogen), \SI{0.23}{\micro\meter} (modal radius) polystyrene (Sigma), and \SI{0.21}{\micro\meter} (modal radius) silica (Kisker). Samples are imaged under flow in straight hydrophilized channels with a height of \SI{20}{\micro\meter} and a width of \SI{800}{\micro\meter} in chips made from Topas (COC, ChipShop).

\subsection{Off-axis Digital Holographic Microscope.} 
A sketch of the setup employed in this study is shown in figure \ref{fig:1}f. A \SI{633}{\nano\meter} HeNe laser (Thorlabs) beam is split into two light beams, one passing through the sample (collected by an Olympus 40X 1.3NA oil objective) and one used as a reference. The two beams are recombined at a slight offset angle, and the resulting interference pattern is recorded by a CCD-camera (AlliedVision, ProSilica GX1920). This interference pattern carries information about the optical field at the camera plane, as described below. Video files of particle samples are recorded at 30 frames per second with typical exposure times in the range of \SIrange{2}{4}{\milli\second}. 

\subsection{Image analysis.}
The interference patterns, or holograms, are analysed using the homemade software written using MATLAB (Mathworks Inc.) to extract the amplitude and phase maps using standard methods\cite{Midtvedt2020}. In brief, due to the off-axis configuration of our setup, the Fourier transform of the interference pattern contains two off-centre peaks, which describe the object field multiplied by a plane wave ($\exp(\pm i \vec{k}_p\cdot \vec{x})$, where $\vec{k}_p$ represents the projection of the wave-vector of the reference beam onto the imaging plane on the camera), in addition to the central peak which corresponds to the non-interferometric intensities. In order to isolate the object field, we numerically shift one of the off-centre peaks to the centre of the Fourier spectrum and apply a low-pass filter. The magnitude and phase of the resulting field correspond to the amplitude and phase of the optical field recorded by the camera. The obtained field is slightly distorted due to optical aberration in the beamline. This is corrected by fitting the phase of the field to a fourth-order polynomial, which is subsequently subtracted from the phase to obtain a close-to-aberration-free image of the real and imaginary part of the optical field. 

\subsection{Particle localisation and tracking.}
In-plane subpixel localisation of detected local extrema in the field amplitude is performed using the radialcenter method\cite{Parthasarathy2012}. Particles are subsequently distinguished from background noise by the degree of radial symmetry and the spatial extent of the local extremum. The z-position of the particle is defined as the plane where the standard deviation of the Fourier transform of the field is minimised\cite{Midtvedt2020}. The scattering pattern of the particle is stored at this plane.  
Following a complete tracking of a frame, each observation is associated with a distance metric to particles in previous frames. Specifically, the observations are joined into traces by minimising the sum of this metric using the Hungarian algorithm\cite{Kuhn1955}. In our setup, the track lengths are typically around 80-100 frames, and all tracks shorter than 20 frames are discarded in the analysis.

\subsection{Particle characterisation network architecture and training.}

The problem of characterising subwavelength particles by their scattering is fundamentally that of noise reduction by averaging multiple observations of the particle and subsequently measuring properties of its scattering pattern, such as the induced phase shift and the radial profile. However, due to small differences in the particle's subpixel position and spatially variable aberrations, observations typically cannot be directly averaged. This motivated the design of WAC-NET (figure \ref{fig:1}g), which consists of two main modules: a representation transforming network, and a regression network. The representation transforming network is a  convolutional neural network followed by a single fully connected layer. It transforms $64\times64\times2$ images centred around a particle, where the two feature channels represent the real and the imaginary part of the complex field, into a one-dimensional vector representation of length 128. From this vector representation, the model also computes a single scalar that represents the confidence in the correctness of the vector representation using two fully connected layers. The regression network, in turn, receives a stack of vector representations and confidence values, representing all observations of a single particle. The confidence values are rescaled using the softmax function. The vector representation stack is averaged along the first axis using the rescaled confidence values as weights. This results in a single vector of length 128. Three fully connected layers then transform this vector into two scalars representing the radius and the refractive index of the particle. 
Both modules are trained simultaneously using synthetic data simulated with Mie theory; the point spread function is calibrated to match the microscope used to collect the data. Standard L1 error is used as loss function, and the Adam optimiser\cite{Kingma2015} is used with a learning rate of 0.001.

\subsection{Simulation of training data.}

The scattered field from dielectric particles is simulated using the MATLAB-package MatScat\cite{Jan-PatrickSchafer2011}. The fields are numerically propagated using the angular spectrum method through a lens with identical NA as the experimental system (NA=1.3) and further propagated to the Fourier plane of the lens. The field at this plane corresponds to the Fourier transform of the field at the camera plane. Next, the optical system is modelled via the pupil function $P(k_x,k_y)$, which relates the simulated fields $F_{\rm sim} (k_y,k_x)$ to the experimentally obtained fields $F(k_x,k_y)$, as $F(\vec{k})=P(\vec{k}) F_{\mathrm sim} (\vec{k})$. This pupil function is estimated based on experimental images of particles of known size and refractive index (polystyrene spheres with radius \SI{0.23}{\micro\meter} and refractive index of  1.58). The pupil function has support only inside a finite radius, i.e. $P(\vec{k})=0$ for $k_x^2+k_y^2>k_{\mathrm P}^2$, where $k_{\mathrm P}$ is set by the numerical aperture of the system as $k_{\mathrm P}=\pi/\lambda\cdot NA$. In our system, $k_{\mathrm P}=$ \SI{6.45}{\per\micro\meter}. This finite support allows considerable dimensionality reduction, as only Fourier components within this radius contain relevant information. This leaves, in our case, 177 pixels in Fourier space. 

\end{methods}



\end{document}